\documentclass[12pt]{iopart}
  
\usepackage{graphicx}
\usepackage{siunitx}
\usepackage{caption, subcaption}
\usepackage{xcolor}
\begin{document}

\title[Broken symmetry state in  P doped BaFe$_2$As$_2$]{Possible signature of broken symmetry state near the quantum critical point in P doped BaFe$_2$As$_2$: A Raman spectroscopy study}

\author{Debasmita Swain$^{1}$, Soumyadeep Ghosh$^{2}$, K. Bera$^{3}$, Sven Friedemann$^{4}$, Haranath Ghosh$^{2}$, Anushree Roy$^{1}$, Sitikantha D Das$^{1}$}

\address{$^1$Department of Physics, Indian Institute of Technology Kharagpur, Kharagpur 721302, India}
\address{$^2$Human Resources Development Section, Raja Ramanna Centre for Advanced Technology, Indore 452013, India; also at Homi Bhabha National Institute (HBNI), Mumbai, India}
\address{$^3$ School of Nanoscience and Technology,  Indian Institute of Technology Kharagpur, Kharagpur 721302, India }
\address{$^4$ Department of Physics, University of Bristol, Bristol, BS8 1QU, United Kingdom}
\ead{sitikantha.das@iitkgp.ac.in}
\vspace{10pt}

\begin{abstract}
	We study the iron-pnictide compound BaFe$_2$(As$_{1-x}$P$_x$)$_2$ for x $\sim$0.23, with a doping concentration near  quantum criticality and enhanced nematic fluctuating state in the  doping-temperature phase diagram. Transport measurements confirm the presence of a magneto-structural transition at 60K from the tetragonal to the orthorhombic phase, followed by a superconducting transition below 16K. The temperature and polarisation dependent Raman spectra reveal that there is a  phonon mode at 211 cm$^{-1}$, followed by two broad modes (BM) between 400 and 700 cm$^{-1}$, having an energy difference of 15 meV, in the temperature range between 300K and 80K. In the non-superconducting state, the phonon mode exhibits expected polarization dependence as well as temperature evolution due to anharmonicity, strong anisotropic and thermally inert behaviour are observed for the BM. Electronic structure calculations for doped and undoped BaFe$_2$As$_2$ show that while Fe $d_{xz}$ and $d_{yz}$ orbitals do not split in the tetragonal phase, the splitting energy is 13.5 meV in the orthorhombic phase of the doped system, which is reasonably close to the experimentally observed value of the energy separation of the BM. We believe that reported BM possibly are the signature of electronic Raman scattering involving the crystal field levels of $d$-orbitals of Fe$^{2+}$ due to local breaking of the C$_4$ symmetry of the parent compound in the doped system.  
\end{abstract}

\section{Introduction}

The iron-based superconductors (FeSCs) can be conveniently classified, depending on their chemical composition and  spacer layers  between the iron pnictogen planes. Among them, the {{\it{A}}Fe$_2$As$_2$} (122 class) is one of the most well studied where superconductivity is induced either by electronic doping \cite{Rotter} or by applied external pressure \cite{Alireza,Takehiro}. The underdoped region of most of the FeSCs has shown crossover between superconducting and spin density wave (SDW) phase and thus suggesting antiferromagnetic spin fluctuation as a strong candidate for superconducting pairing mechanism. In  addition, the underdoped region have shown some intriguing observations related to the electronic phase change; for example, the isovalently substituted compound BaFe$_2$(As$_{1-x}$P$_x$)$_2$ shows strong Fermi to non-Fermi liquid transition at a temperature well above the structural, and magnetic transition and this behaviour is suggested to be due to antiferromagnetic fluctuations \cite{Fermi}. 

The phase diagrams of a large class of FeSCs are well studied by now. The parent high temperature tetragonal phase is ubiquitous to all systems, out of which emerges exotic low temperature phases such as superconductivity and nematicity in the symmetry broken orthorhombic phase. Extensive efforts are made to understand the structure and dynamics of these low temperature phases, which have further led to the experimental evidences of the double Q phase \cite{Frandsen} close to the quantum critical point in a narrow range of doping and temperature. In this scenario, these phases can be thought of as vestigial phases those manifest out of the  parent phase by the successive breaking of  a subset of the tetragonal symmetries \cite{Morten}. It is thus of paramount importance to understand the nature of the high temperature tetragonal phase. 

Several microscopic models are invoked to understand the electronic properties of the pnictide superconductors. The challenge is to find the minimalistic model that will effectively capture all the different phases. Although the puckering of the pnictogen atoms above and below the Fe planes results in a weaker crystal field splitting between the orbitals in the $t_{2g}$ and $e_g$ subsets, yet to understand the low energy properties, it is essentially the $d_{xz}$, $d_{xy}$ and the $d_{yz}$ Fe orbitals that play a major role. These three orbitals are the main contributors to the Fermi surface. Further, as has been shown \cite{Raghu} the effect of the $d_{xy}$ orbital can be integrated out and absorbed in the next nearest neighbour hopping between the $d_{xz}$ and the $d_{yz}$ orbitals thereby making it an effective two orbital model. However, tetragonal symmetry dictates that both $d_{xz}$ and $d_{yz}$ should be degenerate, and thus they should have unequal occupation only in the symmetry broken orthorhombic phase.

Indications that the degeneracy of the  $d_{xz}$ and $d_{yz}$ orbitals being lifted in the tetragonal phase had been borne out earlier. Transport measurements confirmed the presence of finite nematic fluctuations above the structural transition. Angle resolved photoemission spectroscopy (ARPES) studies on some FeSCs reveal that orbital ordering exists well above the structural and magnetic transition temperature for underdoped BaFe$_2$As$_2$, which is suppressed towards the optimally doped region, and almost vanishes in the overdoped region \cite{Yi,Sonobe}. However, since a finite detwinning stress has to be applied for such measurements, the issue of whether the degeneracy between the $d_{xz}$ and $d_{yz}$ orbitals are lifted without the application of any symmetry breaking field still remained. Recent $^{75}$As NMR measurements in LiFeAs and allied 111 compounds have shown that even in the absence of any such external fields, there is in-planar anisotropy of the electric field gradient in the tetragonal phase \cite{Kobayashi}. Structural analysis using X ray pair distribution function measurements on FeSe have also shown that local orthorhombic distortions do exist within the average tetragonal matrix even upto 300K with a concomitant lifting of the $d_{xz}$ and $d_{yz}$ degeneracy \cite{Konstantinova}.

In Raman scattering the observed phonon anomalies  in FeSCs are attributed to phenomena such as first order phase transition, the opening of SDW gap, strong spin-phonon coupling, and change of charge density in FeAs plane \cite{Rahlenbeck}-\cite{Choi}. On the other hand, electronic Raman scattering studies on FeSCs indicate that there exists a superconducting gap \cite{2010,Muschler}, nematic fluctuations in the  tetragonal phase without introducing any external driving force \cite{Gallais}-\cite{Thorsmolle}, and nematic quantum critical point (QCP) near magnetic QCP \cite{Gallais,Thorsmolle,Kretzschmar}. According to a Raman study on the doping dependence of BaFe$_2$(As$_{1-x}$P$_x$)$_2$, a nematic resonance mode has been  observed within the superconducting gap close to nematic QCP \cite{Adachi}. Similar  behaviour has also been encountered for Ba(Fe$_{1-x}$Co$_x$)$_2$As$_2$ \cite{QCP}, suggesting a close relationship between nematic correlations and superconductivity. There are also a few reports on the study of high wavenumber Raman spectra in several FeSCs \cite{2010},\cite{Sugai}-\cite{FeSe1-x}, and the main feature from the analysis was the appearance of wide high energy Raman excitations \cite{2010,Sugai,Pradeep}, which have been attributed to two-magnon excitations. Apart from these wide humps, some high frequency weak and broad Raman modes are observed in Fe based 11 \cite{FeSe,FeSe1-x}, 1111 \cite{Kumar,Zhao}, and 122 \cite{Pradeep} systems, which are claimed to be associated with electronic Raman scattering involving Fe $d$ orbitals.  According to Kumar et al. \cite{Pradeep}, in the case of unstrained crystals of Ca(Fe$_{0.97}$Co$_{0.03}$)$_2$As$_2$, the broad humps centered near $\sim$ 650 cm$^{-1}$ and $\sim$ 800 cm$^{-1}$ correspond to the crystal field splitting involving Fe $d_{xz}$ and $d_{yz}$ orbitals. Similar behaviour has been encountered both theoretically and experimentally for different other FeSCs. It has been argued that this splitting occurs close to structural transition ($T_s$) with splitting energy increasing as the temperature decreases below T$_s$  \cite{Yi,Sonobe,Shimojima,NaFeAs}. Surprisingly, in case of Ca(Fe$_{0.97}$Co$_{0.03}$)$_2$As$_2$ \cite{Pradeep}, there are indications of orbital ordering upto room temperature, yet this observation remained unexplained by Kumar et al. 

Here we study P doped BaFe$_2$As$_2$  because it is one of the least disordered compounds in the representative family of FeSCs. Clear signatures of orbital ordering have been observed in this system well above $T_s$ by ARPES measurements \cite{Sonobe}. Therefore, it is the most suitable system to study orbital anisotropy above $T_s$. To the best of our knowledge, the investigation of high frequency Raman part still remains unexplored in isovalently doped Fe pnictides.

\section{Experimental and computational details}
Single crystal of BaFe$_2$(As$_{1-x}$P$_x$)$_2$ with $x\sim$0.23 was synthesized by the self-flux method. Micro-Raman spectroscopy in the back-scattering geometry was carried out using a  triple monochromator (T64000, Horiba, France) in the subtractive mode. The spectrometer is equipped with an optical microscope (BX41, Olympus, Japan) along with an objective lens 50$\times$L ((numerical aperture 0.50), a Peltier cooled CCD detector (Synapse, Horiba, France), and a water-cooled Argon-Krypton ion laser (Innova 70C Spectrum, Coherent, USA).  Spectra were recorded using 514 nm laser excitation wavelength. The heating effect of the laser power on the sample was checked and avoided in our study. In our measurements, the laser power on the sample was kept $<$ 1 mW.  Temperature-dependent Raman measurements were carried out using a sample stage (THMS-600, Linkam, UK) over the temperature range between 80K and 300K.

All electronic structure and phonon mode calculations has been performed using plane wave pseudo-potential based method as implemented in the QUANTUM ESPRESSO package \cite{Giannozzi}.
 We use generalized gradient approximation (GGA) of Perdew-Burke-Ernzerhof (PBE) for the exchange correlation functional \cite{Burke}. 
 The experimental crystal structure reported in Ref. \cite{Marianne} 
 for the tetragonal phase with space group symmetry I4/mmm and orthorhombic phase with space group symmetry Fmmm are  used for non-spin polarized single point energy calculations. The Ultrasoft pseudopotentials are taken from Psilibrary database. Virtual Crystal Approximation (VCA) is employed to dope the system theoretically \cite{Bellaiche}.
  In VCA method, fictitious virtual atoms replaced the actual atoms in same atomic site. The Kohn-Sham orbitals are expanded in plane wave basis set and kinetic energy cutoff is set to 65 Ry for undoped sample after performing rigorous convergence test. Charge density cutoff is set to 520 Ry. We use Monkhorst-Pack scheme \cite{Monkhorst,Pack} 
  for Brillouin zone (BZ) sampling in $k$-space, and the grid size used for self-consistent field (SCF) calculation is 12$\times$12$\times$12, also chosen after careful convergence test. Broyden, Fletcher, Goldfrab, and Shanno (BFGS) \cite{Broyden}
  geometry optimization scheme is used to yield optimized structures, where the force and total energy convergence threshold are set to 10$^{-5}$ and 10$^{-8}$, respectively (in a.u.).
  For phonon calculation, 4×4×4 $q$-mesh is used for both systems with Gaussian broadening method along with the Gaussian width of 0.02 Ry. Zone centered ($q$ = 0, 0, 0) phonon spectra (Raman modes) are determined by using density functional perturbation theory (DFPT) \cite{Baroni}.  
  
  \section{Results}
  \begin{figure}[h]
  	\centering
  	\includegraphics[scale=0.7]{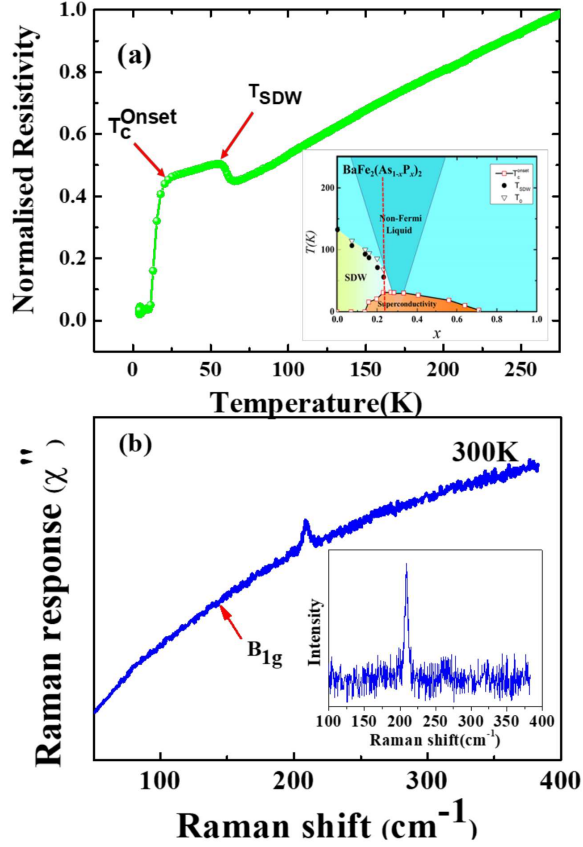}
  	\caption{(Color by electronic media) (a) Temperature dependent normalised resistivity  upto 275K. 
  		Inset: Schematic phase diagram of P doped BaFe$_2$As$_2$ compound reproduced from Ref.\cite{Fermi}.(b) Raman response within a spectral range of 50-400 cm$^{-1}$ at 300K. 
  		Inset: Unpolarised Raman spectrum at 300K.}
  	\label{resistivity}
  \end{figure}

We have characterised BaFe$_2$(As$_{1-x}$P$_x$)$_2$ ($x$ $\approx$ 0.23), with electrical transport measurements. Fig. 1(a) shows in-plane normalised resistivity of the compound below 275K. 
The resistivity anomaly observed around 60K is attributed to a magneto-structural SDW transition.
The resistivity shows a sharp drop at a lower temperature with zero resistivity attained below 10K. The superconducting transition temperature corresponding to our resistivity data lies within the  doping range of 0.20 to 0.23, whereas the SDW transition temperature corresponds to a doping composition of $x\approx$ 0.23, as can be seen from the inset phase diagram of Fig 1(a), which is reproduced from Ref \cite{Fermi}. 

Fig. 1(b) shows the Raman response $\chi^{''}$ of BaFe$_2$(As$_{0.77}$P$_{0.23}$)$_2$, where,
$\chi^{''}$= $I(\omega)/1+n(\omega)$,  $I(\omega)$ is the Raman intensity and  $1+n(\omega)$= $1/(1-e^{-\hbar\omega/k_{B}T})$ is the thermal Bose factor with $k_B$ as the Boltzmann constant.  In a recent report \cite{Adachi}, the same spectral feature (having B$_{1g}$ symmetry) of BaFe$_2$(As$_{1-x}$P$_x$)$_2$ has been studied in detail to establish a correlation between the superconductivity and the nematic fluctuation in this compound. The inset of Fig. 1(b) plots the room-temperature Raman spectrum of BaFe$_2$(As$_{0.77}$P$_{0.23}$)$_2$ over the spectral range between 100 and 375 cm$^{-1}$. In this range, only one Raman mode at 211 cm$^{-1}$ is observed. 

From the group theory analysis of the tetragonal I4/mmm space group, the irreducible representation of the $\Gamma$ point phonon modes are $\Gamma$=$A_{1g}$+$B_{1g}$+2$E_g$. We have calculated the phonon spectra (Raman modes) at the Gamma point ($q$ = 0, 0, 0) of BaFe$_2$(As$_{1-x}$P$_{x}$)$_2$ ($x$ $\approx$ 0.20) using first-principles based DFPT, as described in Section 2. To the best of our knowledge, the lattice constants of the given doped compound are not available in the literature. Thus, in our simulation, we used (i) experimental lattice constants of BaFe$_2$As$_2$ in its tetragonal phase as reported in ref. \cite{Marianne}, and (ii) 
theoretically determined lattice constants of BaFe$_2$(As$_{1-x}$P$_x$)$_2$ with $x$ $\approx$ 0.20 for the calculation of the phonon modes. We further calculated Raman modes of the parent compound using (i). Experimentally observed Raman peak at 215 cm$^{-1}$ with B1g symmetry \cite{Baum} appers at 219 cm$^{-1}$ in our theoretical calculation. 

Following the methodology in (i), the calculated A$_{1g}$, B$_{1g}$, and 2E$_g$ modes of the doped compound are obtained at 193.0, 213.5, 133.4, and 237.7 cm$^{-1}$, respectively. The same by following the methodology (ii) are estimated as 206.3, 225.8, 141.4 and 294.2 cm$^{-1}$.  While the A$_{1g}$ mode corresponds to vibrations of the As ions, B$_{1g}$ mode represents Fe-ion vibrations along the z-axis. $E_g$ modes originate from Fe- As- vibrations in the ab-plane with respect to Ba spacer. The experimentally observed  Raman peak at 211 cm$^{-1}$ corresponds to B$_{1g}$ mode  at 213.5 cm$^{-1}$. By following the methodology (ii) (with  the theoretically determined lattice constants), it appears at 225.8 cm$^{-1}$ in the theoretical prediction.  This may be due to the error involved in the theoretical estimates of low frequency Raman modes using GGA-PBE. A similar kind of disagreement with theoretical lattice constants is also found in the case of another doped FeSC compound CeFeAsO$_{1-x}$F$_x$ \cite{Kumar}. It is to be noted that first principles calculation indicates that all phonon modes in this compound lie below 300 cm$^{-1}$.

\begin{figure}
	\centering
	\includegraphics[scale=0.7]{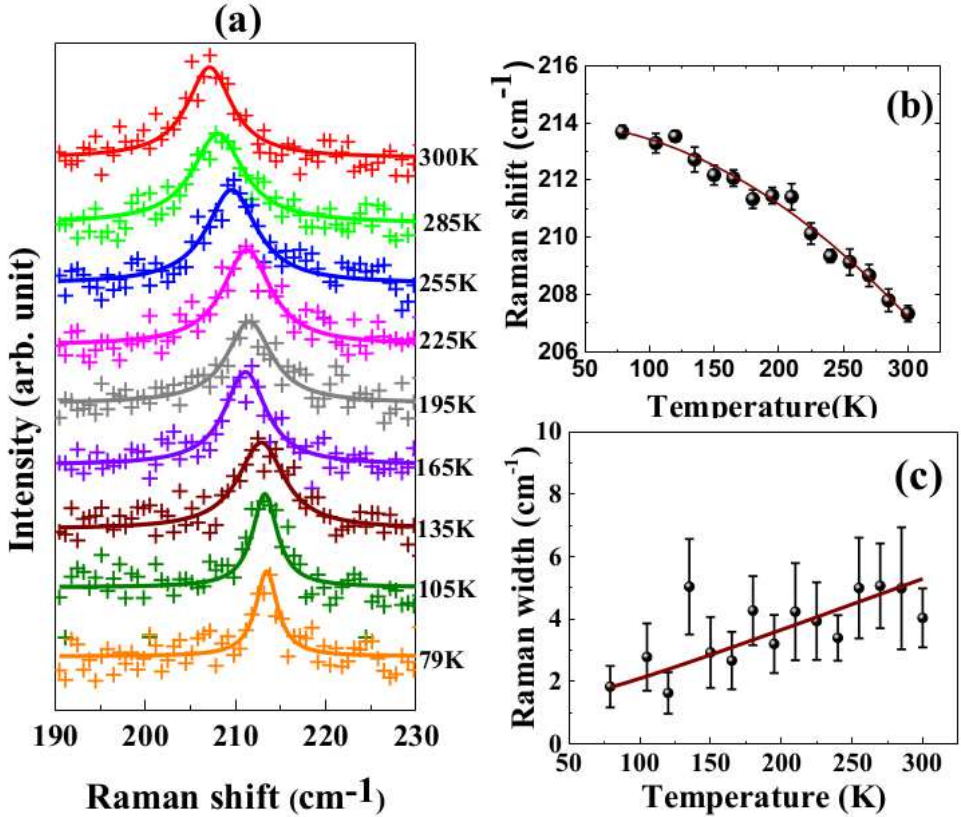}
	\caption{(Color by electronic media) (a) Characteristic Raman spectra (+ symbols) of BaFe$_2$(As$_{0.77}$P$_{0.23}$)$_2$ at various temperatures, mentioned in the right of the panel. The solid lines are the best fit to the data points with a Lorentzian profile.  Temperature dependence of (b) the Raman shift and (c) Raman linewidth of the B$_{1g}$ phonon mode. The error bars are the standard deviation of the parameters as obtained from the fitting procedure. The solid lines are the best fit to the data points as described in the text.}
	\label{Raman2}
\end{figure}

Fig. 2(a) plots characteristic Raman spectra (+ symbols) of BaFe$_2$(As$_{0.77}$P$_{0.23}$)$_2$ over the spectral window between 190 and 230 cm$^{-1}$ at various temperatures between 80K and 300K. Following the phase diagram, shown in Fig. 1(a), the system is expected to be in the tetragonal normal state over this temperature range. Each spectrum was fitted with a Lorentzian function and shown by the solid lines. Fig. 2(b) and 2(c) show the temperature dependence of the Raman shift and Raman width (full width at half maxima), respectively, as obtained from the fitting procedure. The error bars are the standard deviation of the fitted parameters.  Various phenomena can contribute to the shift in the Raman spectral profile with the temperature of a system. The shift ($\Delta\omega_j$) in phonon wavenumber ($\omega_j$) due to lattice expansion with the increase in temperature is given by the relation  $\frac{\Delta \omega_j}{\omega_j}$=$\gamma\frac{\Delta V}{V}$ where $\Delta V$ is the change in unit cell volume V. $\gamma$ is the dimensionless Gr\"{u}neisen parameter. By considering the reported \cite{Ni} change in volume by 0.5 $\si{\angstrom}^3$ (from 399.9
\si{\angstrom}$^3$ to 400.4 \si{\angstrom}$^3$) over the temperature 
range between 80K and 300 K of BaFe$_2$As$_2$ and the value of $\gamma$ in the normal state \cite{Fujii}, we estimate $\Delta\omega$ of the given mode over the temperature range between 80K and 300K as 3-6$\times$10$^{-3}$ cm$^{-1}$, which is much smaller than that observed Raman shift in Fig. 2(b). Therefore, the thermal expansion alone cannot be responsible for the observed shift in the peak position. The shift in phonon wavenumber  with the increase in anharmonicity can be estimated from the relation \cite{Balkanski}:
\begin{equation}
\centering
\omega(T)=\omega_o + A\left( 1+\frac{2}{e^{\hbar\omega_o/2K_BT}-1}\right) + B\left( 1+\frac{3}{e^{\hbar\omega_o/3K_BT}-1}\right), 
\end{equation}
where $\omega_o$  is the Raman wavenumber at $T$=0 and $A$ and $B$ are  anharmonic constants.
The data points in Fig. 2(a) are fitted with Eqn. 1 (see solid red line). The fitting parameters are $\omega_o$ = 214 cm$^{-1}$, $A$= -0.262 cm$^{-1}$, and $B$= -0.099 cm$^{-1}$.
The thermal evolution of Raman line-width of the same mode at 211 cm$^{-1}$ is shown in Fig. 2(c). As in the case of Raman shift, due to anharmonicity, the Raman linewidth ($\Gamma$) is expected to vary with temperature by following the relation \cite{Balkanski}. 
\begin{equation}
\centering
\Gamma(T)= C\left( 1+\frac{2}{e^{\hbar\omega_o/2K_BT}-1}\right) + D\left( 1+\frac{3}{e^{\hbar\omega_o/3K_BT}-1} \right). 
\end{equation}
Here $C$ and $D$ are anharmonic constants. The solid red line is the best fit to the data points using Eqn. 2  with the fitting parameters are $C$=1.36 cm$^{-1}$
and $D$= -0.002 cm$^{-1}$. 
The above results indicate that the B$_{1g}$ mode at 211 cm$^{-1}$ behaves as a normal phonon mode over the temperature range between 80K and 300K.
\begin{figure}
	\centering
	\includegraphics[scale=0.8]{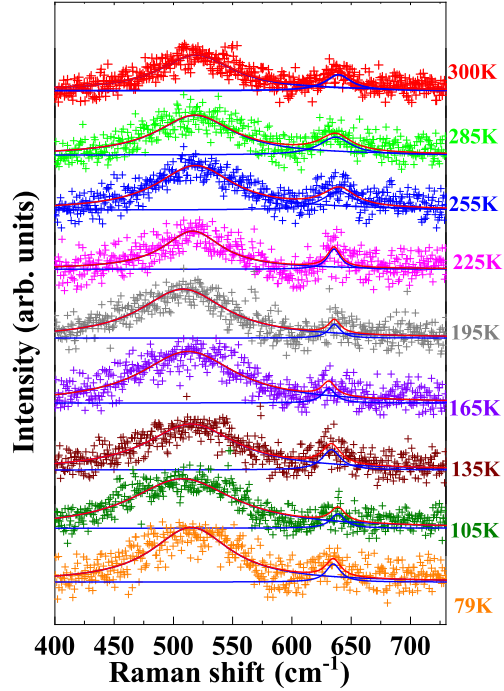}
	\caption{(Color by electronic media)  Characteristic unpolarized Raman spectra (+ symbols) over the spectral range between 400 and 725 cm$^{-1}$. Blue solid lines are fit to the individual BM, and solid red lines are the net fits to the data. }
	\label{Raman3}
\end{figure}

In addition to the $B_{1g}$ mode, shown in Fig. 2, we have observed two other weak broad modes (BM) over the spectral range between 400 and 725 cm$^{-1}$. A few characteristic Raman spectra, recorded at different temperatures, over this spectral range are shown in Fig \ref{Raman3}. It is to be noted that in comparison to the Raman mode at 211 cm$^{-1}$, these spectral features are broad. Each spectrum was fitted with two Lorentzian functions (shown by solid blue lines). The net fitted spectrum is shown by the red lines. The Raman shifts of these BM appear at 515 and 635 cm$^{-1}$. Importantly,  within the error bars of the fitting parameters, we do not observe any significant change in Raman shift, and width of these broad bands with temperature, unlike the observed B$_{1g}$ mode. The separation between the broad spectral features remains at $\sim$15 meV for all temperatures. We observe a decrease in the intensity of the band at 515 cm$^{-1}$ peak with temperature (shown in Fig. 4(a)). Because of the low signal-to-noise ratio, it is non-trivial to comment on the thermal evolution of the intensity of the other peak at 635 cm$^{-1}$ or the spectral widths and peak positions of these broad bands with temperature.
\begin{figure}
	\centering
	\includegraphics[scale=0.6]{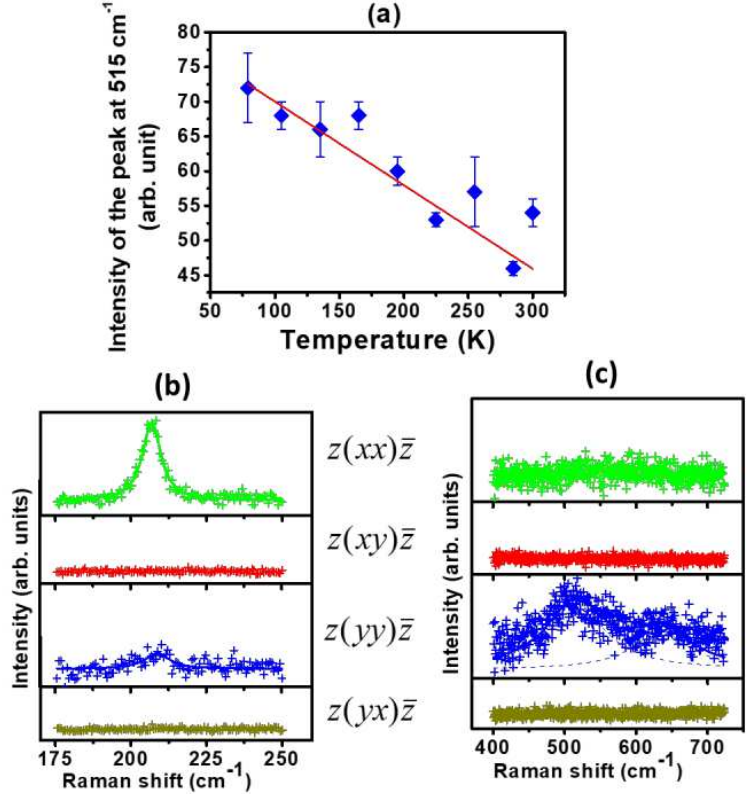}
	\caption{(Color by electronic media)
		(a) The  variation of the intensity of the peak at 515 cm$^{-1}$ with temperature. (b) Polarization dependent Raman spectra of the broad modes over the  range between 400 and 725 cm$^{-1}$ following Porto's notation.}
	\label{Raman4}
\end{figure}
\\

Raman spectra of the B$_{1g}$ mode at 220 cm$^{-1}$ and the  double hump at 515 and 625 cm$^{-1}$ in  Z(XX)\={Z}, Z(XY)\={Z}, Z(YX)\={Z} and Z(YY)\={Z}  scattering geometries, are shown in Fig. 4(b) and (c). As mentioned earlier, the B1g mode is polarized along the c axis of the crystal. A weak signal in  Z(YY)\={Z}  configuration may be due to the polarization leakage from the imperfect crystal orientation and/or scattering geometry. Z is along the $c$ axis of the crystal, so that XY represent the $ab$ plane of the crystal. From Fig. 4(c) it appears that the broad bands are polarized in the $ab$ plane of the crystal. Thus, it is  most likely that they carry the signature of the in-plane Fe ion or Fe-As ions in the lattice.
\\

In the present case, unlike the observed thermal evolution of the B$_{1g}$ phonon peak at 211 cm$^{-1}$, the two BMs at 535 and 615 cm$^{-1}$ respectively do not exhibit any significant temperature dependence over the temperature range between 300K and 80K. Interestingly, these modes do have a strong polarisation dependence, where the maxima of the intensity in the scattered channel corresponds to the direction of polarisation of the incident radiation  along the Fe-Fe bond angle. As the population of a ground state decreases with temperature, the Raman cross-section associated with the crystal field excitation is expected to drop with the rise in temperature, as shown in Fig. 4(a). Signatures of such electron Raman scattering are often weak \cite{FeSe, Kumar}.

\section{Discussion}

\begin{figure}[h!]
	\centering
	\includegraphics[scale=0.4]{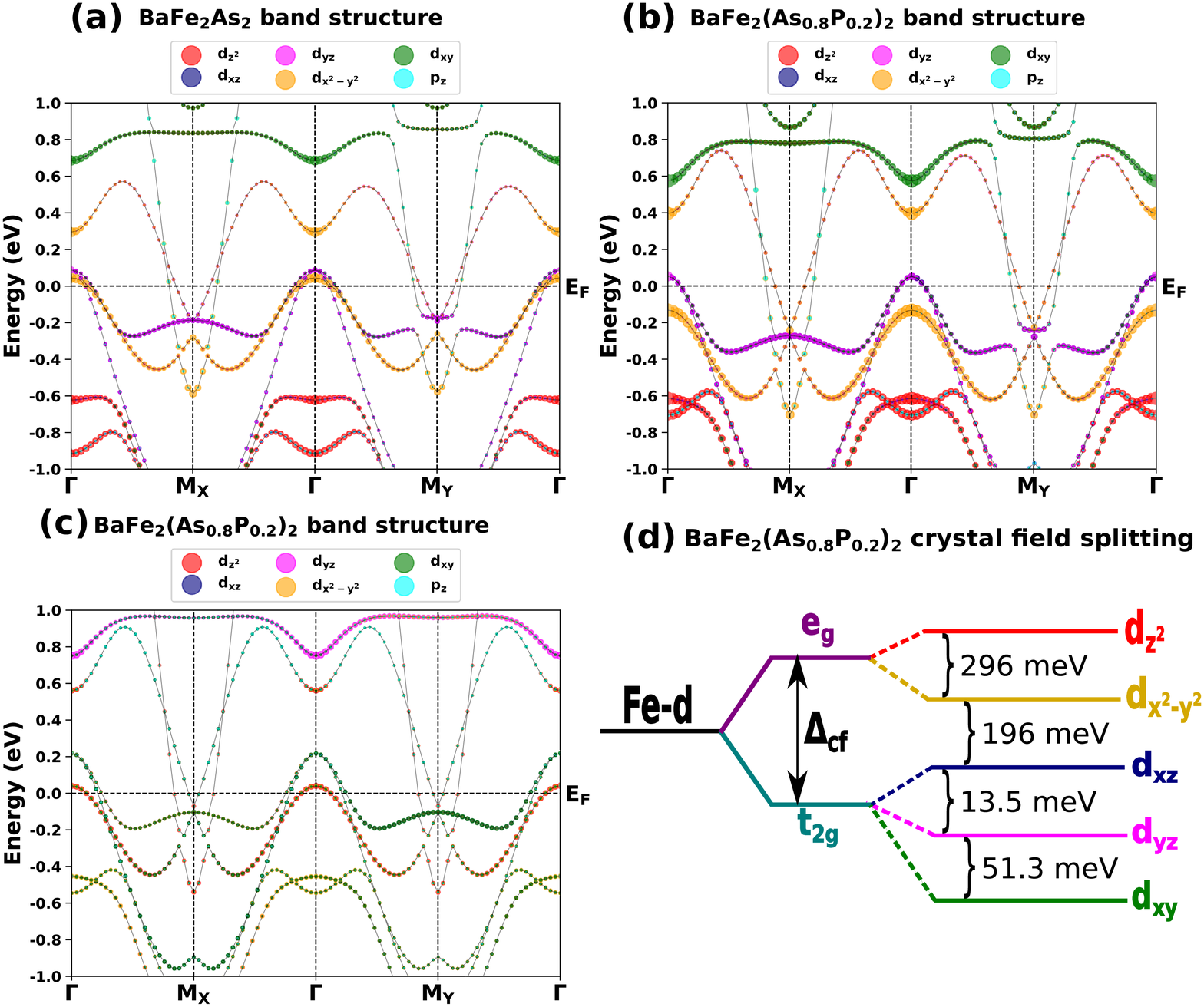}
	\caption{(Color by electronic media)(a)Theoretically calculated orbital projected band structures of different Fe-3$d$ and As-4$p_z$ orbitals for BaFe$_2$As$_2$, (b) BaFe$_2$(As$_{0.8}$P$_{0.2}$)$_2$ at tetragonal phase, and (c) BaFe$_2$(As$_{0.8}$P$_{0.2}$)$_2$ at orthorhombic phase around high symmetry points. Different colors show individual orbital contributions, and orbital weight is proportional to the sizes of the circles. The Fermi energy level is set to zero. (d) Crystal field splitting of BaFe$_2$(As$_{0.8}$P$_{0.2}$)$_2$ at orthorhombic phase. Energy levels in meV indicate the difference between different $d$-orbitals.}
	\label{5}
\end{figure}

\begin{figure}[h!]
	\centering
	\includegraphics[scale=0.65]{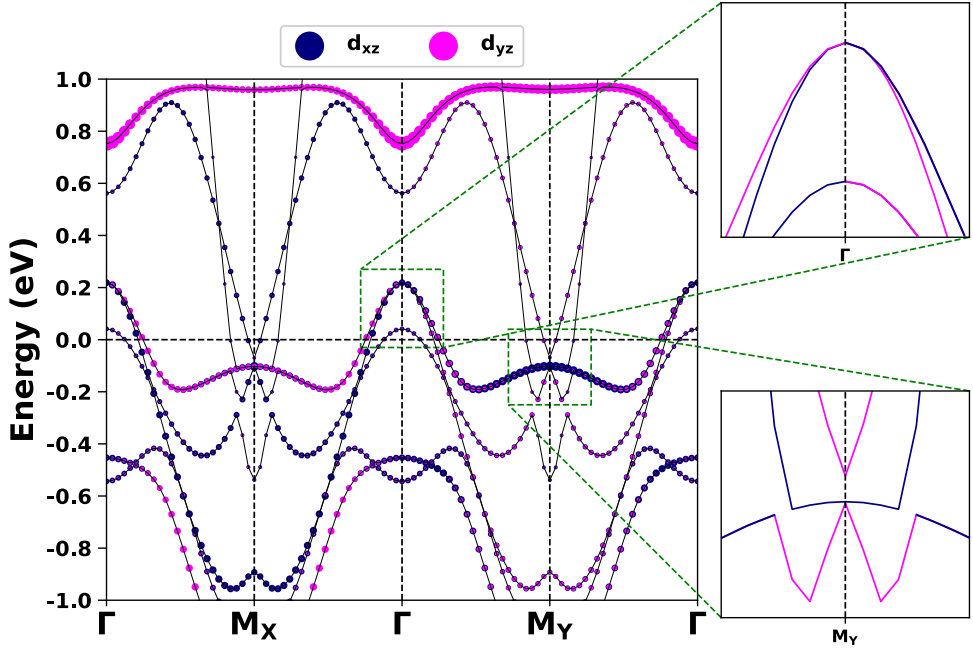}
	\caption{(Color by electronic media) Electronic band structure of BaFe$_2$(As$_{0.8}$P$_{0.2}$)$_2$ showing momentum dependent  $d_{xz/yz}$ band splitting along $\Gamma$-$M_{x(y)}$ directions}
	\label{6}
\end{figure}

To decipher the origin of the BM, we carried out the orbital projected electronic band structure of the parent and doped compounds in the tetragonal phase and presented it in Fig. \ref{5}. Different orbital contributions which are involved in the band formation are denoted by assigning a colour code. The size of each circle is proportional to the weight of the orbitals. Red, magenta, green, blue, yellow, and cyan colors are assigned to signify Fe-$d_{z^2}$, $d_{yz}$,  $d_{xy}$,  $d_{xz}$,  $d_{x^2-y^2}$, and As-$4p_z$ orbitals, respectively. 
The undoped BaFe$_2$As$_2$ compound possesses three hole-like bands around the $\Gamma$-point and two electron-like bands around X-points ($M_X$ and $M_Y$) \cite{Singh}. The doped sample contains two-electron like bands around X-points and two hole-like bands around the $\Gamma$-point. The orbital projected electronic band structure reveals mixed multi-band multi-orbital nature of both the compounds. For the undoped compound, there are three hole-like-bands crossing the Fermi level (FL). Among the three bands two are almost degenerate and mostly have mixed $d_{xy}$ and $d_{yz}$ character. The third band (near to FL) is primarily of $d_{x^2-y^2}$ and $d_{yz}$ orbitals derived. There are two electron-like bands around X-points crossing the FL. The electron like band close to the FL is primarily $d_{z^2}$ orbital derived. The other electron-like bands have mixed $d_{yz}$, $d_{z^2}$ characters, ref. Fig \ref{5}(a). In the case of the tetragonal phase of doped sample two hole-like bands are degenerate and have mixed $d_{xy}$ and $d_{yz}$ character. The band with $d_{x^2-y^2}$ orbital contribution lies below the FL due to doping. Hence, these two nearly degenerate upper lying bands are at the verge of Lifshitz transition \cite{Ghosh1,Ghosh2}. The lowest energy hole-like band around X-point has mixed $d_{x^2-y^2}$ and $d_{z^2}$ orbital character, while the other hole-like band has a dominant contribution from $d_{yz}$ orbitals along with a miniature contribution from the As $4p_z$ orbitals, refer to Fig. \ref{5}(b). However, the electronic band structure does not reveal any band splitting to describe the broad bands observed in Fig. 3.

We have also simulated the band structure of the doped compound in the orthorhombic phase. In this case, the hole-like band lies above the FL is primarily of $d_{xy}$ and As 4$p_z$ orbitals derived. Another hole-like band lies below the FL has contributions from $d_{z^2}$ and d$_{x^2-y^2}$ orbitals. There are two electron-like bands around X-point crossing the FL. The electron-like band close to the FL is primarily of $d_{z^2}$ and As $4p_z$ orbital derived. The other electron-like band is of mixed $d_{xy}$, $d_{z^2}$ characters(See Fig. \ref{5}(c)). Fig. \ref{5}(d) shows the crystal field splitting of BaFe$_2$(As$_{0.8}$P$_{0.2}$)$_2$ at the orthorhombic phase. The crystal field splitting has been calculated using maximally localized Wannier function based formalism \cite{Scaramucci} as implemented  in the Wannier90 package\cite{Scaramucci}. We get an effective low energy model by constructing ten maximally localized Wannier functions freezing states in finite energy window above and below the Fermi level (-1.0 to 1.0 eV) \cite{Ghosh1}. The onsite energy difference between five $d$-orbitals are termed as crystal filed splitting \cite{Scaramucci}. The crystal field splitting is a finger print of atomic like states whereas the electronic band structures are the real electronic states of solid in momentum space. Momentum dependent of $B_{1g}$ modes are such that it picks up the in-plane electronic structures along  $\Gamma$-$M_{X(Y)}$ direction. Thus observation of  the temperature indpendent broad modes between 400 and 700 cm$^{-1}$ are consistent with $d_{xz/yz}$ splitted bands of Fe. 

High frequency broad modes  have been observed in other FeSCs viz. FeSe$_{0.82}$ \cite{FeSe} and CeFeAsO$_{1-x}$F${_x}$ \cite{Zhao}.  These spectral features are attributed to crystal field splitting, which survives upto 300K. X-ray pair distribution function measurements on FeSe \cite{Konstantinova} demonstrate that local orthorhombicity survives well above the structural transition upto room temperature. Furthermore, NMR measurements on P doped BaFe$_2$As$_2$ have indicated the presence of inplane electric field gradients even in the tetragonal phase with a corresponding lifting of 3$d_{xz/yz}$ degeneracy \cite{Tetsuya}.  Electronic band structures were obtained from first principle calculations for 0.23\% P doped BaFe$_2$As$_2$ as well as the  undoped system in the tetragonal phase, which show that $d_{x^2-y^2}$ undergoes a Lifshitz transition while the $d_{xz/yz}$ is on the verge of such a transition in the doped compound. However, it is hard to rationalise a scenario within this framework where the orbital degeneracy between the $d_{xz}$ and the $d_{yz}$ orbitals are lifted. Since Raman scattering is a local probe it perhaps could `see' the local symmetry breaking either in the form of orthorhombic distortion or lifting of orbital degeneracy even at 300K, which is well over the structural transition of this system. We believe that the broad humps arise due to the electronic Raman scattering involving the d-orbitals of Fe in this compound as observed in temperature-dependent Raman study of a CeFeAsO$_{0.9}$F$_{0.1}$ superconductor \cite{Kumar}.

\section{Summary}
We have presented temperature dependent  Raman measurements on BaFe$_2$(As$_{1-x}$P$_x$)$_2$, which is established as one of the less disordered compounds in the iron-pnictide family. The study is on the tetragonal phase of the compound i.e. the temperature range was chosen well above the expected magnetic or structural phase transitions of the system. Other than the normal temperature variation of the B$_{1g}$ phonon mode, we discuss the appearance of the broad double humps in high wavenumber regime. The invariant spectral profile with temperature suggests that the origin of these broad peaks is electronic in origin. Furthermore, we observe a strong polarization dependence of these peaks. The orbital projected first principles electronic band structure calculation suggests Lifshitz transition of the   $d_{x^2-y^2}$ orbital in the tetragonal phase. However, no band splitting could be observed. We attribute the observed electronic transition to the consequence of local symmetry breaking, as indicated in the literature in the related compound.

\end{document}